\documentclass[epsf]{aipproc}

\layoutstyle{6x9}

\def\Journal#1#2#3#4{{#1} {\bf #2}, #3 (#4)}

\def\PRL{\em Phys. Rev. Lett.}

\def\beq{\begin{equation}}
\def\eeq{\end{equation}}

\def\be{\begin{equation}}
\def\ee{\end{equation}}
\def\bea{\begin{eqnarray}}
\def\eea{\end{eqnarray}}
\def\ba{\begin{array}}
\def\ea{\end{array}}
\def\dis{\displaystyle}
\newcommand{\End}{\end{document}}

\def\D0{D\O~}

\def\f{\frac}

\def\to{\rightarrow}

\def\tanb{\tan\hspace*{-1mm}\beta}

\def\ts{\tilde{t}}
\def\cs{\tilde{c}}

\def\({\left(}
\def\){\right)}
\def\[{\left[}
\def\]{\right]}

\def\sq2{\sqrt{2}}

\def\tanb{\tan\hspace*{-1mm}\beta}

\def\U1em{{U(1)_{\rm em}}}

\def\sq2{\sqrt{2}}

\def\tanb{\tan\hspace*{-1mm}\beta}
\def\ee{e^+e^-}

\def\ms{{\widetilde{m}}}
\def\sm0{{\widetilde{m}_0}}

\def\U1em{{U(1)_{\rm em}}}

\def\sq2{\sqrt{2}}

\def\tanb{\tan\hspace*{-1mm}\beta}

\newcommand{\lae}{\stackrel{<}{\sim}}

\def\dis{\displaystyle}

\renewcommand{\thefootnote}{\fnsymbol{footnote}}

\begin{document}

\title{Higgs: Standard Model and Beyond\footnote{
Talk given at VIII Mexican Workshop on Particles and Fields, 
Zacatecas, Mexico, November 14-20, 2001.
}}


\author{C.-P. Yuan}{
address={Department of Physics and Astronomy,
East Lansing, MI 48824, USA},
email={yuan@pa.msu.edu},
}



\begin{abstract}
In this talk, I will discuss possible new physics effects that
modify the interaction of Higgs boson(s) with top and bottom
quarks, and discuss how to detect such effects in current and
future high energy colliders.
\end{abstract}

\maketitle

\setcounter{footnote}{0}
\renewcommand{\thefootnote}{\arabic{footnote}}

\section{Introduction}
Two of the great mysteries in the elementary particle physics are
the cause of the electroweak symmetry breaking (that generates
masses for the weak gauge bosons $W^\pm$ and $Z$) and the origin
of the flavor symmetry breaking (that generates masses for quarks
and leptons). In the Standard Model (SM), both symmetry breaking
mechanisms are explained by introducing a single Higgs boson
doublet field. The $W$ and $Z$ bosons gain their masses from
Goldstone boson mechanism and the fermions gain their massed from
Yukawa interactions. In the SM, the mass of the Higgs boson can
receive a large radiative correction that is proportional to
the square of the cutoff scale beyond which new physics effect has
to take place. In order for the SM to be a valid field theory all
the way up to the Planck's scale (about $10^{19}$\,GeV), the mass
of the SM Higgs boson has to be somewhere around 130\,GeV to
180\,GeV (to satisfy the naturalness condition). Furthermore, to
explain the diverse fermion mass spectrum, every fermion has to be
assigned a different Yukawa coupling. The fact that the mass of
the top quark is so large as compared to the other fermions and is
close to the vacuum expectation value suggests that top is a {\it
special} quark and it may play a role in the electroweak symmetry
breaking. If that is the case, then the flavor symmetry breaking
and the electroweak symmetry breaking mechanisms may be related.
In this talk I will discuss two different classes of models -- one
is the strongly interacting models and another is the weakly
interacting models. (For a brief review on the SM Higgs
boson, please see the talk by J. Conway in this Workshop.)
The typical strongly interacting models are
Technicolor, top-condensate~\cite{topCrev}, topcolor~\cite{Hill}
 and top-seesaw models~\cite{he_hill}, and
the weakly interacting models are supersymmetry models,
particularly,
the minimal supersymmetric standard model (MSSM)~\cite{MSSM}.
In the former class of models, the electroweak
symmetry breaking is generated dynamically and usually results in
composite (in contrast to elementary) Higgs bosons.
On the contrary, in the latter class of models, the electroweak
symmetry breaking is generated spontaneously and results in
elementary Higgs bosons.
I shall take the topcolor model and the MSSM as two examples to discuss
the phenomenology predicted by these two classes of models and
to identify a few experiments at high energy colliders that
can distinguish these models assuming some new physics signals are
found.

\section{Models}
\subsection{Topcolor model}
In the topcolor model, the mass of the top quark is generated by
topcolor dynamics that also contributes to the electroweak
symmetry breaking and induces two composite Higgs boson doublets
in its low energy effective theory. The physical (composite)
scalars are $t$-Higgs~($H_t^0$), top-pions~($\pi_t^0,~\pi_t^\pm$)
and $b$-Higgs~($H_b^0$,~$A_b^0,~H_b^\pm$). Because the topcolor
dynamics has to be strong enough to make top quark and antiquark
to form condensate to generate the large top quark mass as well as
to contribute to part of the weak boson masses, the Yukawa
couplings of $t$-Higgs~ (or top-pions) with top quarks have to be
large (at the order of 1). Furthermore, because the bottom quark
is the isospin partner of the top quark, the strong topcolor
dynamics that a left-handed top quark experiences will also affect
the bottom quark, Hence, the Yukawa coupling of $b$-Higgs~ with
bottom quarks have to be large as well. This should be compared
with the SM in which the Yukawa coupling of top quark to the SM
Higgs boson is about 1 while the coupling of bottom quark is much
less than 1 (about 1/50 at the 100 GeV scale). In this model, tau
lepton does not involve the topcolor dynamics so that it does not
directly couple to the composite scalars and its Yukawa coupling
vanishes. (Its mass has to be generated by other mechanics, such
as the technicolor dynamics. Again, this is different from
the SM prediction which is equal to $\sqrt{2} m_\tau / v$ where
$v$ is the vacuum expectation value ($\sim 246$\,GeV) and $m_\tau$
is the mass of tau. Hence, it is expected that the collider
phenomenology of this model will be significantly different from
the SM.

As noted above, there are also charged Higgs bosons and top-pions
predicted in this model. Their couplings to the top, bottom, and charm
quarks are shown in the equation below:
\beq
\ba{l}
{\cal L}_{\pi_t}^{tc} \hspace*{-1.1mm}=\hspace*{-1.1mm}
\dis\f{m_t\tanb}{v}\hspace*{-1.1mm}\left[
iK_{UR}^{tt}{K_{UL}^{tt}}^{\hspace*{-1.3mm}\ast}\overline{t_L}t_R\pi_t^0
\hspace*{-1mm}+\hspace*{-1.2mm}\sq2
{K_{UR}^{tt}}^{\hspace*{-1.3mm}\ast}K_{DL}^{bb}\overline{t_R}b_L\pi_t^+
+ \right.
\nonumber\\[2.3mm]
~~~~\left.
i{K_{UR}^{tc}}{K_{UL}^{tt}}^{\hspace*{-1.3mm}\ast}
\overline{t_L}c_R\pi_t^0
\hspace*{-1mm}+\hspace*{-1.2mm}\sq2
{{K_{UR}^{tc}}}^{\hspace*{-1.3mm}\ast}K_{DL}^{bb}\overline{c_R}b_L\pi_t^+
\hspace*{-1mm}+\hspace*{-1mm}{\rm h.c.}       \right],
\ea
\label{eq:Ltoppi}
\eeq
where {\small $\tanb = \sqrt{(v/v_t)^2-1}$} and
$v_t\simeq O(60-100)$\,GeV is the top-pion decay constant;
$K_{UL,R}$ and $K_{DL,R}$ are rotation matrices that
diagonalize the up- and down-quark mass matrices
$M_U$ and $M_D$, i.e.,
{\small $~K_{UL}^\dag M_U K_{UR} = M_U^{\rm dia}~$} and
{\small $~K_{DL}^\dag M_D K_{DR} = M_D^{\rm dia}$},~
from which the CKM matrix is defined as
{\small $~V=K_{UL}^\dag K_{DL}~$}.
As shown in Ref.\,\cite{hy}, a typical topcolor model,
that is consistent with all the precision low energy data,
 gives
\beq K_{UR}^{tt}\simeq
0.99\hspace*{-0.5mm}-\hspace*{-0.5mm}0.94~,~~~ K_{UR}^{tc}\lae
0.11\hspace*{-0.5mm}-\hspace*{-0.5mm}0.33~,~~~ K_{UL}^{tt} \simeq
K_{DL}^{bb} \simeq 1 ~. 
\label{eq:KURtc} \eeq 
As to be discussed
later, the large flavor mixing between $t_R$ and $c_R$ can lead to
very distinct collider signatures. One example is to induce a
large single-top event rate at hadron colliders~\cite{tim}.

\subsection{MSSM}
In the MSSM, the electroweak symmetry is radiatively broken due to
the contribution of the heavy top quark in loops. (It would not
have worked if the top quark were not heavy enough.) Therefore, in
this model, top quark also plays a special role in the electroweak
symmetry breaking. Two Higgs doublet fields are required in the
MSSM by the requirement of supersymmetry. Among the eight real
fields, three of them are the Goldstone bosons which generate the
masses of the weak gauge bosons and five of them are the two
CP-even Higgs bosons ($h$ and $H$), one CP-odd Higgs boson ($A$)
and two charged Higgs bosons ($H^\pm$). Their couplings to the
fermions are derived by demanding one Higgs doublet couple to the
up-type fermions and another to the down-type fermions, which is
similar to a type-II two-Higgs-doublet model~\cite{MSSM}. For
example, the tree level Yukawa couplings of $A$-$b$-$b$ and
$A$-$t$-$t$ are
 $\sqrt{2} m_b \tan\beta / v$, and
 $\sqrt{2} m_t \cot\beta / v$, respectively,
 where
$\tan\beta$ is the ratio of the two vacuum expectation values.
For a large $\tan \beta$, the coupling of $A$-$b$-$b$
can become $O(1)$, while the coupling of $A$-$t$-$t$
becomes very small.
This pattern of the Yukawa couplings
is not only different from that in the SM (where the
top Yukawa coupling is
much larger than the bottom Yukawa coupling)
but also different from the topcolor model
(where both the top and bottom Yukawa couplings are $O(1)$).
This difference is the crucial element that allows us to
distinguish different classes of electroweak symmetry braking models
by carefully examining the experimental data.
We shall come back to this point in the next section.

A perfect supersymmetric theory cannot describe the Nature.
(Otherwise, we would have seen various supersymmetric partners of the
observed particles.) Hence, supersymmetry has to be broken.
To incorporate the effect from
 the yet-to-be found supersymmetry breaking mechanism,
the MSSM contains the soft-breaking sector in its Lagrangian to
parameterize all such possibilities. One interesting possibility
induced by a general soft-breaking sector is that the top-squark
and the charm-squark can be largely mixed to yield a sizable
flavor mixing between top and charm.
One such model can be generated by imposing a
horizontal $U(1)_H$ symmetry, which is called {the  Type-B}
supersymmetry model in Ref.~\cite{dhy}. Another way to generate
the up-type squark mixings is to have a non-diagonal $A_u$ in the
flavor space. Motivated by the charge-color-breaking (CCB) and
vacuum stability (VS) bounds, we define at the weak scale
$$
A_u' =
      \left(
      \ba{ccc}
      0 & 0 & 0\\
      0 & 0 & x\\
      0 & y & 1
      \ea
      \right) A \,,
$$
where {$(x,\,y) = O(1)$} and
$A_u'$ is $A_u$ in the
super-CKM basis (where the quark mass matrix
{$M_u$} is diagonal).
Hence, we have generated large flavor-mixings in the
$\ts - \cs$ sector, which are consistent with
all low energy experimental flavor changing neutral current
(FCNC) data
and theoretical {CCB/VS} bounds.
Without losing generality, we define
the Type-A1 model as $x\neq 0,~y=0$ and
Type-A12 model as $x=0,~y\neq 0$.
It is obviously that in the former model
$\tilde{c}_L$ decouples, and in the latter model
$\tilde{c}_R$ decouples.
(Here, we assume
$
M_{LL}^2 \,\simeq\, M_{RR}^2\, \simeq\,\sm0^2\,{\bf I}_{3\times3}
$,
 for simplicity.)
It has been shown in Ref.~\cite{dhy} that
 both Type-A and Type-B models can
radiatively generate large flavor-mixing Yukawa couplings to
quarks. 

\section{$gg,q {\bar q} \to b {\bar b} \phi^0$~~ and
$b {\bar b} \to \phi^0$ } In Ref.~\cite{he_bot}, we calculated the
cross sections of the $gg,q {\bar q} \to b {\bar b} \phi^0$
processes at the Tevatron and the LHC, where $\phi^0$ denotes a
(pseudo-) scalar predicted in the strongly interacting topcolor
model, the two Higgs doublet extension of top-condensate model
\cite{tt-2HDM}, and the MSSM. To suppress the large SM QCD
backgrounds in the detection mode of $\phi^0 \to b {\bar b}$, a
set of kinematic cuts has to be applied together with 3 or more
$b$-tags. For the class of
strongly interacting models, the Tevatron Run~II and the LHC
are able to either exclude an entire model or a large part of the
model parameters if the $\phi^0b {\bar b}$ signal is not found
experimentally. Likewise, for the class of weakly interacting
models, such as the MSSM, a large portion of the supersymmetry
parameters on the $\tan\beta$ versus $m_A$ plane can be probed at
the Tevatron and the LHC. However, because the coupling of
$A$-$b$-$b$ can receive large (about a factor of 2) radiative
corrections from the supersymmetric particle threshold effect and
the QCD interaction, the precise region of $\tan\beta$ as a
function of $m_A$ that can be studied via the above processes will
depend on other supersymmetry parameters, such as the $\mu$
parameter and the top-squark masses~\cite{he_bot}.
 Nevertheless, the constraint
provided by this process on the $\tan\beta - m_A$ plane is
complementary to that provided by the associated production of the
weak gauge boson and the Higgs boson.

If the mass of $\phi^0$ is large, it can be dominantly produced from the
s-channel fusion process $b {\bar b} \to \phi^0$ at hadron colliders.
The cross sections of this process for various models were also
given in Ref.~\cite{ch_qcd}.

As noted in the previous section, to distinguish the strongly
interacting from the weakly interacting models, one should also
examine the production of $t {\bar t} \phi^0$ in addition to the
$b {\bar b} \phi^0$ channel. Furthermore, to distinguish those two
classes of models in the $b {\bar b} \phi^0$ channel, one can examine
the tau lepton decay mode of $\phi^0$. (In the topcolor model,
$b$-Higgs~do not couple strongly to the tau lepton, while in the
MSSM with a large $\tan\beta$, the coupling of
$\phi^0$-$\tau$-$\tau$ is large.) 

\section{$c {\bar b} \to H^+$}
If the mass of a charged Higgs boson is
lees than $\sim 170$\,GeV, it can be studied via
the decay process $t \to H^+ b$
using the large data sample of the $t \bar t$ pairs
expected at the Tevatron and the LHC~\cite{tbH}.
For a heavy charged Higgs boson, the $H^+H^-$ pair production
rate is usually small unless enhanced by some resonant effect.
One such example is to have have a heavy neutral Higgs boson
with mass larger than twice of $m_{H^+}$ in the MSSM.
It can also be associated produced with a top quark via
$g b \to H^\pm t$ \cite{gbHt}, but again with a small rate.
In~\cite{dhy,hy}, it was pointed out that
$H^+$ can be produced via the s-channel process
$cs \to H^\pm$ because of the large parton luminosities of
charm and strange quarks at the LHC.
Furthermore, if the
flavor-mixing coupling of $c$-$b$-$H^+$ can be large, then
$H^+$ can also be produced via  $cb \to H^\pm$.

In the topcolor model, the mass of top-pion $\pi_t^\pm$ is expected to
be around the weak scale.
For a given $m_{\pi_t}$, the allowed range of the
Yukawa couplings has to be checked by comparing with low energy
precision data. A recent study for the topcolor assisted
technicolor model can be found in Ref.~\cite{kuang}. To test this
model's prediction, we can study the single-top event signature
from $ c {\bar b} \to H^+ \to t {\bar b}$. The invariant mass
distribution of the $t$-$\bar b$ system can reveal the existence
of such a resonant~\cite{ch_qcd}. To truly test this
model, one should also check that the polarization of the final
state top quark is right-handed because of its couplngs, cf.
Eq.~(\ref{eq:Ltoppi}). In contrast, the top quark produced from
the SM single top processes, either the s-channel process $q {\bar
q'} \to W^\ast \to t {\bar b}$ or the t-channel process $q b \to
q' t$, are almost one hundred percent left-handedly 
polarized~\cite{tim}.

\begin{figure}
\resizebox{0.5\textwidth}{!}
{\includegraphics{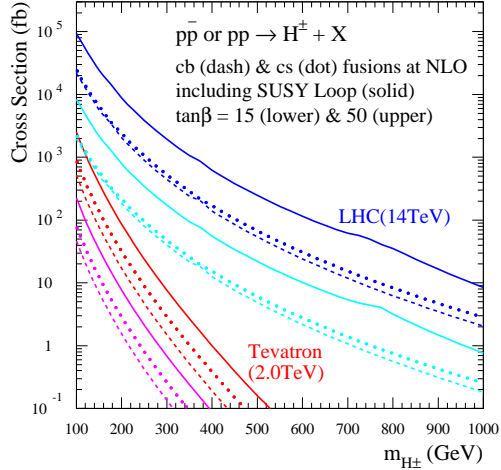}}
\caption{$H^\pm$ production via $cb$ (and $cs$) fusions at
hadron colliders.}
\label{Fig:Sigma_mssm}
\end{figure}

Similarly, in the MSSM, a sizable flavor-mixing $c$-$b$-$H^+$
coupling can be radiatively generated through radiative correction
arising from the large stop and scharm mixings. At the tree level,
the coupling of $c$-$b$-$H^+$ is suppressed by the CKM matrix
element $V_{bc}$ which is about 0.04. Hence, even with the large
enhancement factor from a large $\tan\beta$, the tree level rate
of $c {\bar b} \to H^+$ is still smaller than that of $c {\bar s} \to
H^+$ at the Tevatron and the LHC because the parton luminosity of
the strange quark is much larger than that of the bottom quark. As
shown in Ref.~\cite{dhy}, the Type-A supersymmetry models with the
non-diagonal scalar trilinear $A$-term for the up-type squarks can
enhance the $c$-$b$-$H^+$ coupling from the contribution of stop,
scharm and gluino in loops. For $y=0$ and $x \sim O(1)$ (i.e.
Type-A1 model), the production rate of $c {\bar b} \to H^+$ can be
increased by a factor of 2 to 5 as compared to its tree level
rate, depending on the value of $x$. In Fig.~\ref{Fig:Sigma_mssm},
we show the single charged Higgs boson production rate for a
typical choice of the supersymmetry parameters
$(m_{\tilde{g}},\mu,\ms0))= (300,\,300,\,600)$\,GeV,
$(A,\,-A_b)=1.5$\,TeV, and $\tanb=(15,\,50)$ with $x=0.75$.

\section{$t \to c \phi^0$}
The flavor-mixing dynamics predicted by models can also be tested from
the FCNC decay of the top quark, such as
$t \to c \phi^0, c \gamma, c g$.

It is known that the SM branching ratio of the flavor-changing top
decay $t \to c h^0$ is extremely small
($\lae 10^{-13}-10^{-14}$ \cite{SM-tch}), so that
this process provides an excellent window
for probing new physics.
In the topcolor model The low energy precision data requires the
mass of the top-pions not  to be too small as compared to the top
quark mass~\cite{kuang}. In case that $m_{\pi}^0 < m_t -m_c$, the
decay process $ t \to c \phi^0$ can occur at tree level,
cf. Eq.~(\ref{eq:Ltoppi}), which can impose further constraint on the model
if such a signal is not found experimentally. A few recent studies
on the other FCNC processes predicted by the topcolor model can be
found in Ref.~\cite{yue}.

In the MSSM, the loop induced $t$-$c$-$h^0$ coupling can be
greatly enhanced depending on the detailed parameters of the
model.~\cite{dhy}. Assuming that the only dominant
decay mode of the top quark is its SM decay mode, i.e. $t \to bW$,
the decay branching ratio of $t \to c h^0$ is 
given by  
Br$[t\to ch^0]\simeq
\Gamma [t \to ch^0]/\Gamma [t \to bW]$. As summarized in
Table~\ref{Tab:fsu2}, ${\rm Br}[t\to ch^0]$ can be as large as
$10^{-3} - 10^{-5}$ over a large part of the supersymmetry
parameter space where the mass of the lightest Higgs boson $h^0$
is around $110-130$\,GeV. Since the LHC with an integrated
luminosity of $100$\,fb$^{-1}$ can produce about $10^8$ 
$t \bar{t}$ pairs, it can have a great sensitivity to discover this
decay channel and test the model predictions, by demanding one top
decaying into the usual $bW^\pm$ mode and another to the FCNC
$ch^0$ mode.

\begin{table}[h]
\vspace*{5mm}
\caption{
Br$[\,t\!\to\! c\,h^0\,]\times 10^{3}$
is shown for a sample set of Type-A1 inputs with
$(\sm0,\mu,A)=(0.6,0.3,1.5)$\,TeV and Higgs mass
$M_{A^0}=0.6$\,TeV.
The three numbers in each entry correspond to
$x = (0.5,\,0.75,\,0.9)$, respectively.
}
\vspace*{1.5mm}
\begin{tabular}{c||c|c|c}
\hline
&&&\\[-2.5mm]
$m_{\tilde g}$ & $\tanb=5$ & 20 & 50~         \\ [1.5mm]
\hline
&&&\\[-2.5mm]
\,$100$\,GeV   & (.011,\,.10,\,.81)  &  (.015,\,.19,\,4.6)
             & (.016,\,.21,\,7.0)\,   \\[1.5mm]
\hline
&&&\\[-2.5mm]
\,$500$\,GeV &  (.011,\,.09,\,.41)  &   (.015,\,.13,\,1.0)
           &  (.016,\,.14,\,1.2)\,   \\[1.5mm]
\hline
\end{tabular}
\label{Tab:fsu2}
\end{table}

\section{Conclusion}
Because the mass of the top quark is close to the weak scale, it may
play an essential role in the breaking of the electroweak symmetry.
Two classes of models -- strongly interacting and weakly interacting
models --- are considered.
In the topcolor model, the Yukawa couplings of the top quark 
are large. As an isospin partner of the top quark, the bottom quark 
can also experience large Yukawa interactions.
With the possibility of having a large flavor-mixing between the 
(right-handed) top- and charm-quarks,  
 the charged top-pions can be copiously produced via the s-channel
 $cb$-fusion process at high energy colliders, 
 due to its large Yukawa coupling and the sizable 
 parton luminosities.
In the MSSM, the flavor symmetry is tightly connected to the
supersymmetry breaking through the introduction of the soft
breaking sector in the Lagrangian. To carefully study the flavor-mixing
and the flavor changing neutral current
 processes can advance our knowledge on the supersymmetry breaking
 mechanism. This point was demonstrated in some production and decay 
 processes. 
 Although through out this talk, I only concentrated on the
 phenomenology at hadron colliders, some similar effects are also
 expected in the future Linear Colliders. For example, a polarized
 $\gamma \gamma$ collide can test the chirality of the Yukawa coupling
 of $b$-$c$-$H^+$ by studying the single charged Higgs boson
 production~\cite{kanemura}.
There are two other aspects of the physics of Higgs boson which I do not 
have space to write on this report. 
First, the associated production
of $A$ and $H^\pm$ from the quark fusion process can constrain MSSM as a 
function of the only one supersymmetry parameter $m_{A}$, because 
the coupling of $W^\mp$-$A$-$H^\pm$ is determined by the gauge interaction
and $M_{A}$ and $M_{H^+}$ are related by $M_{W}$ in the MSSM \cite{ky}.
This mass relation however does not generally hold in a 2HDM.
Second, in a general 2HDM, the CP-odd scalar can be very light so that 
it can register into detectors as a "single-photon" signature 
(which originates from a pair of highly boosted photons 
produced by the decay of $A$). In that case, the production 
process $q {\bar q'} \to W^\pm h$ with $h \to A A$ 
and $A \to \gamma \gamma$ appear as a   
 $W^\pm + \gamma \gamma$ signal event at colliders~\cite{gilberto}.

\section*{Acknowledgments}
I would like to thank J.L Diaz-Cruz and M.A. Perez 
for their support, and my collaborators C. Balazs, L.J. Diaz-Cruz,
H.-J. He, S. Kanemura, F. Larios, G. Tavares-Velasco, and T. Tait
 for their invaluable contributions.
This work was supported in part by the NSF grant PHY-9802564.


\end{document}